\documentclass[aps,prc,preprintnumbers]{revtex4}

\usepackage{amssymb,amsmath}
\usepackage{epsfig}
\usepackage{bm}

\newcommand{\vect}[1]{\bm{#1}}  

\newcommand{\cc}{$c\bar{c}$}

\newcommand{\ssbar}{$s\bar{s}$}

\newcommand{\be}{\begin{equation}}
\newcommand{\ee}{\end{equation}}
\newcommand{\bea}{\begin{eqnarray}}
\newcommand{\eea}{\end{eqnarray}}
\newcommand{\bean}{\begin{eqnarray}}
\newcommand{\eean}{\end{eqnarray*}}

\newcommand{\gapproxeq}{\lower
.7ex\hbox{$\;\stackrel{\textstyle >}{\sim}\;$}}
\newcommand{\lapproxeq}{\lower
.7ex\hbox{$\;\stackrel{\textstyle <}{\sim}\;$}}

\def\3bar{$\bar {\hbox{\bf 3}}$}
\newcommand{\qq}{$q\bar{q}$}
\newcommand{\ket}[1]{| {#1} \rangle}
\newcommand{\bra}[1]{\langle {#1} |}       
\newcommand{\rme}[3]{\bra{#1}|{#2}|\ket{#3}}
\newcommand{\me}[3]{\bra{#1}{#2}\ket{#3}}
\newcommand{\ot}{\otimes}
\newcommand{\st}[1]{\Pi_{#1}}
\newcommand{\sixj}[6]{ \left\{\begin{array}{ccc}  {#1} & {#2} & {#3} \\    {#4} 
& {#5} & {#6} \end{array}   \right\} }
\newcommand{\ninej}[9]{ \left\{\begin{array}{ccc}  {#1} & {#2} & {#3} \\    {#4} 
& {#5} & {#6}\\    {#7} & {#8} & {#9}  \end{array}   \right\} }

\begin{document}

\title{Looking for a gift of Nature: Hadron loops and hybrid mixing}
\author{F. E. Close}
\email[E-mail: ]{f.close1@physics.ox.ac.uk}
\affiliation{Rudolf Peierls Centre for Theoretical Physics, University of Oxford,\\ 1 Keble Road, Oxford, OX1 3NP, UK}

\author{C. E. Thomas}
\email[E-mail: ]{thomasc@jlab.org}
\affiliation{Thomas Jefferson National Accelerator Facility, Suite \#1, 12000 Jefferson Avenue, Newport News, Virginia 23606, USA}

\date{13 March 2009}
\preprint{OUTP-09-01P}
\preprint{JLAB-THY-09-934}

\begin{abstract}
We investigate how coupling of valence \qq~ to meson pairs can modify the properties of conventional \qq~ and hybrid mesons. In a symmetry limit the mixing between hybrids and conventional \qq~ with the same $J^{PC}$ is shown to vanish.  Flavor mixing between heavy and light \qq~ due to meson loops is shown to be dual to the results of gluon mediated pQCD, and qualitatively different from mixing involving light flavors alone. The validity of the OZI rule for conventional \qq~ and hybrid mesons is discussed.
\end{abstract}

\maketitle

\section{Introduction}
\label{sec:Introduction}

If hybrid mesons or glueballs are ever to be identified as states that are distinct from conventional \qq~ with the
same $J^{PC}$, or from the hadronic continuum, some gift of nature may be required. To illustrate what 
we mean, and to motivate the subsequent discussion, recall 
that the underlying valence \qq~ structure to the light flavored mesons was historically identified 
because nature left some
multiplets relatively clean. The $1^{--}$ and $2^{++}$ multiplets are to good approximation flavor 
eigenstates, the I=0 members
being almost pure $n\bar{n} \equiv (u\bar{u} + d\bar{d})/\sqrt{2}$ and \ssbar~ respectively. Had this 
not been the case, it is
probable that the quark model for mesons would have been delayed until the subsequent discovery of 
another gift, namely
the occurrence of heavy flavors
and associated charmonium or bottomonium spectroscopies.

Glueballs and hybrids are expected to decay into channels that are also accessed by the decays
of conventional mesons. Even when the gluonic hadrons have exotic $J^{PC}$, they can couple to 
pairs of mesons sharing these quantum numbers. In general therefore one may expect that diagrams 
involving meson loops
will cause mixing between
gluonic hadrons and conventional \qq. One of the questions that we shall study here is
if there are circumstances where hybrid states, in particular, may decouple from conventional \qq~ with the same $J^{PC}$. 
If such were to occur, this could be the sought for ``gift of nature" enabling a clean hybrid signal 
to be identified.

A further question is whether there are $J^{PC}$ multiplets for hybrids which are expected to be ideal 
flavor eigenstates. 
In such cases, the correlations of mass and flavor throughout a nonet can distinguish between a hybrid 
nonet and a
tetraquark or di-meson system: the former has the canonical degenerate I = 0 (isospin singlet) and I=1 (isospin triplet) low-lying with a 
single I=0 (\ssbar) above
the strange members (as in the conventional $1^{--}$), whereas the latter has an inverted structure with the degenerate I=0 and I=1 heaviest, above the strange and a low-lying isolated I=0 state, as seen for the scalar
mesons below 1 GeV\cite{PDG08,jaffe,ct00}.

These questions have a common feature: under what circumstances do hadron properties deduced from 
``valence" \qq~ eigenstates avoid large corrections due to hadron loops?
For example, what protects $\omega$ and $\phi$ from mixing through their common coupling to 
$K\bar{K}$?
Discussion of the latter has a long history, as summarised for example in Refs.\ \cite{NAT,lipkin,geiger}.  We shall assess the implications of these works for the flavor mixing in hybrid nonets. We find that
flavor mixing for exotic $J^{PC}$ is likely to be suppressed but that the states are unlikely to be as 
pure flavor eigenstates
as those of the conventional \qq~ nonets with $J^{PC} = 1^{--}$ or $2^{++}$.

Recently some general theorems have been developed\cite{bs08} regarding the effects of hadron loops on
quark model predictions\cite{NAT}. These theorems build on the factorisation of hadron and constituent 
spins in strong interaction vertices that are OZI-allowed\cite{bct}, and become exact in a particular 
mass-degenerate limit. The theorems assumed that the creation of \qq~, which triggered the OZI-allowed decay, is
in spin-triplet\cite{bct,bs08}; we show the theorem holds for either spin-triplet or spin-singlet.
The generalisation confirms the result of Ref.\ \cite{bs08} that mixing vanishes between \qq~ with 
different orbital
quantum numbers, such as $^3S_1$ and $^3D_1$, and we show also that within the same hypothesis,
there is no mixing between these states and hybrid vector mesons.

These theorems will be violated by single gluon exchange. This has particular significance when heavy 
flavors are involved\cite{cd08}. We show that flavor mixing of \cc~ into the $\eta$ or $\omega$ is 
qualitatively different from the \ssbar-\qq~ mixing of light flavors in those systems. In particular, 
the charmed hadron loops connecting $\psi-\omega$ or $\eta-\eta_c$ are dual to the mixing driven by 
gluon intermediate states in pQCD.

\section{Meson Loops}
\label{sec:Loops}

Non-perturbative hadron loops, which are required by unitarity, can play an important role in shifting hadron masses and mixing higher Fock states into \qq~ wavefunctions.  They may also mix different \qq~ configurations of the same $J^{PC}$. The formalism underlying this was developed long ago and an extensive set of applications made in a series of papers (Refs.\ \cite{NAT,Tornqvist:1995kr} and references therein).  Inter alia Ref.\ \cite{NAT} noted that the contribution of a loop of hadrons spanning an $SU(6)_W$ ($SU(2)$ spin $\ot$ $SU(3)$ flavour) multiplet could give sum rules whereby states in a flavor octet would be shifted uniformly in mass.  When applied to charmonium, the masses of $\psi$ and $\eta_c$ were found numerically to be shifted by a similar amount\cite{Heikkila:1983wd}.  In addition, for the $\chi_{cJ}$ states a similar size of mass shift was found for all $J$\cite{Ono:1983rd}.

Recently this numerical phenomenon has been rediscovered\cite{bs08}. The theoretical underpinning of these results has been identified and theorems developed\cite{bs08} building on the factorisation of hadron and constituent spins in strong interaction vertices\cite{bct}. In particular these theorems showed that vector mesons in $^3S_1$ and $^3D_1$ do not mix in a certain symmetry limit.  The purpose of the present paper is to build on these theorems, strengthen their validity and extend them to new situations.  The salient features of Refs.\ \cite{bs08} and \cite{bct} which form the point of departure for the present paper are as follows.

In Ref.\ \cite{bct} it was shown that OZI amplitudes can be factored into contributions depending on 
the total $J$, constituent 
spins $S$ and an overall spatial dependence including the angular momentum of the outgoing partial 
wave. The structure of the 
factored amplitude for the valence-continuum coupling in the particular case of \qq~ creation in 
spin-triplet was written in 
Ref.\ \cite{bct}.  For a meson $A$ (orbital angular momentum $L_A$, spin $S_A$ and total angular 
momentum $J_A$) decaying to 
mesons $B$ ($L_B$, $S_B$, $J_B$) and $C$ ($L_C$, $S_C$, $J_C$) with $J_{BC} = J_B \ot J_C$ and the 
final mesons in partial 
wave $L$, the result was:
\begin{multline}
\me{(L\ot((L_B \ot S_B)_{J_B}\ot(L_C \ot S_C)_{J_C})_{J_{BC}})_{J_A}}{H}{(L_A\ot S_A)_{J_A}} = 
\\
\sum_{S_{BC} L_{BC} L_f}(-)^{L+L_{BC}+L_A+S_A+S_B}\st{X L_{BC}S_{BC}S_{BC}J_BJ_CJ_{BC}L_fS_AS_BS_C}
\\
\sixj{L}{L_{BC}}{L_f}{S_{BC}}{J_A}{J_{BC}}
\sixj{L_f}{L_A}{X}{S_A}{S_{BC}}{J_A}
\ninej{L_B}{S_B}{J_B}{L_C}{S_C}{J_C}{L_{BC}}{S_{BC}}{J_{BC}}
\ninej{1/2}{1/2}{S_B}{1/2}{1/2}{S_C}{S_A}{X}{S_{BC}}
\rme{(L \ot (L_B \ot L_C)_{L_{BC}})_{L_f}}{\vect{\psi}}{L_A}
\label{factored}
\end{multline}
with $\st{a b ...} \equiv \sqrt{(2a+1)(2b+1)...}$.  In the $^3P_0$ model where the \qq~ pair is 
created in 
spin-triplet: $H=\vect{\sigma}\cdot\vect{\psi}$ where $\vect{\sigma}$ is a vector in spin space, $X=1$ 
and $\vect{\psi}$ acts
 on the spatial (orbital and radial) degrees of freedom.  If instead the \qq~ pair is created in 
spin-singlet: $H=\sigma \psi$ 
 with $\sigma$ a scalar in spin space, $X=0$ and $\psi$ again acts on the spatial degrees of freedom. 

Note that in the above expression it is the spatial matrix element $\rme{(L \ot (L_B \ot 
L_C)_{L_{BC}})_{L_f}}{\vect{\psi}}{L_A}$ 
that enforces parity conservation.

\begin{figure}[ht]
\includegraphics[width=9cm]{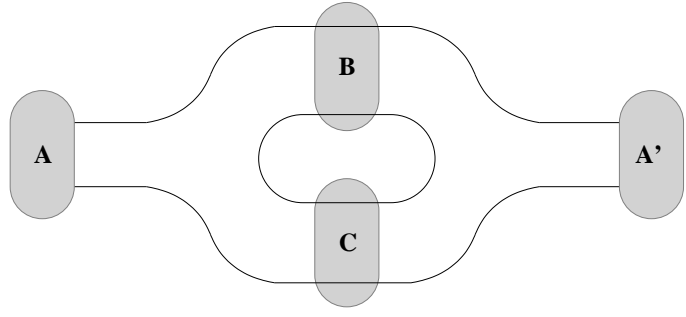}
\caption{Mixing via hadronic loops}
\label{fig:LoopsMixing}
\end{figure}

The mixing amplitude $a_{A' A}$ between initial $A$ and final $A'$
meson valence \qq~ states when summed over a loop of degenerate mesons, labelled $BC$, 
is related to the mass shift $\Delta m(A)$.  The diagram for mixing via loops is shown in 
Fig.\ \ref{fig:LoopsMixing}: this consists of two decays of the same topology sewed together.  
Denoting 
\begin{equation}
\Psi \equiv \int \frac{d^3p}{(2\pi)^3} \frac{1}{m_A - E_{BC}(p) +i\epsilon}
\label{energyintegral}
\end{equation}
as the common energy integral that may be taken outside the sums for intermediate states $BC$ with 
identical masses, then
\begin{equation}
\Delta m(A) = \Psi \sum_{BC} \langle (L_A \ot S_A)_{J_A}| H |BC \rangle \langle BC| H |(L_A \ot 
S_A)_{J_A} \rangle
\label{mass1}
\end{equation}
and
\begin{equation}
(m_A - m_{A'})a_{A'A} = \Psi \sum_{BC} \langle (L_{A'} \ot S_{A'})_{J_{A'}}| H |BC \rangle \langle BC| H | 
(L_{A} \ot S_{A})_{J_{A}} \rangle
\label{mix1}
\end{equation}

Substituting the decay amplitudes from Eq.\ \ref{factored} into Eq.\ \ref{mix1} with $J_{A'}=J_A$ 
gives:
\begin{multline}
(m_A - m_{A'})a_{A' A} = \Psi \sum_{BC}\sum_{S_{BC} L_{BC} L_f}\sum_{S'_{BC} L'_{BC} 
L'_f}(-)^{L_A+S_A+L_{A'}+S_{A'}+L_{BC}+L'_{BC}} \\
\st{X X L_{BC} L'_{BC} S_{BC} S_{BC} S'_{BC} S'_{BC} J_B J_B J_C J_C J_{BC} J_{BC} L_f L'_f S_A S_{A'} 
S_B S_B S_C S_C}
\\
\sixj{L}{L_{BC}}{L_f}{S_{BC}}{J_A}{J_{BC}}
\sixj{L}{L'_{BC}}{L'_f}{S'_{BC}}{J_A}{J_{BC}}
\sixj{L_f}{L_A}{X}{S_A}{S_{BC}}{J_A}
\sixj{L'_f}{L_{A'}}{X}{S_{A'}}{S'_{BC}}{J_A}
\\
\ninej{L_B}{S_B}{J_B}{L_C}{S_C}{J_C}{L_{BC}}{S_{BC}}{J_{BC}}
\ninej{L_B}{S_B}{J_B}{L_C}{S_C}{J_C}{L'_{BC}}{S'_{BC}}{J_{BC}}
\ninej{1/2}{1/2}{S_B}{1/2}{1/2}{S_C}{S_A}{X}{S_{BC}}
\ninej{1/2}{1/2}{S_B}{1/2}{1/2}{S_C}{S_{A'}}{X}{S'_{BC}}
\\
\rme{(L \ot (L_B \ot L_C)_{L_{BC}})_{L_f}}{\vect{\psi}}{L_A, n_{A}}
\rme{(L \ot (L_B \ot L_C)_{L'_{BC}})_{L'_f}}{\vect{\psi}}{L_{A'}, n_{A'}}^{\dagger}
\label{equ:mixfullexpression}
\end{multline}
where $n_{A}$ and $n_{A'}$ represent any other quantum numbers, such as radial excition, of states $A$ and $A'$ respectively.
The orthonormality of 6-j and 9-j symbols in Eq.\ \ref{equ:mixfullexpression} were shown in Ref.\ \cite{bs08} 
to place significant constraints for mixings or energy shifts arising from the coupling 
between valence \qq~ and meson continuum states.  In particular, if $BC$ are OZI-allowed mesons, then 
while individual loop contributions \qq~ $\to B C \to$ \qq~ may give 
large contributions to physical observables, simple closure relations were found when a degenerate 
subset of mesons was summed over.  In particular Ref.\ \cite{bs08} found that the masses of all 
$\chi_{cJ}$ states are shifted by the same amount when the meson intermediate states 
$D\bar{D}; D\bar{D^*}; D^*\bar{D}; D^*\bar{D^*}$ are summed over, in the limit where $M(D) = M(D^*)$. This 
explains the numerical results\cite{Ono:1983rd,bs08}. The theorem also explained the numerical result\cite{Heikkila:1983wd,bs08} 
that $\Delta m(\psi) \approx \Delta m(\eta_c)$ and $\Delta m(\psi') \approx \Delta m(\eta'_c)$.

The result in Eq.\ (A.7) of \cite{bs08} from summing over what we shall call a \emph{semi-complete} 
set of states 
(namely a set of degenerate states with all possible allowed $S_B$, $S_C$, $J_B$, $J_C$ and $J_{BC}$ 
for fixed $L$, $L_B$ and $L_C$) is:
\begin{multline}
(m_A - m_{A'})a_{A' A} = \frac{1}{2 L_A + 1} \delta_{L_A, L_{A'}} \delta_{S_A, S_{A'}} \delta_{J_A, J_{A'}} \\
\sum_{L_{BC} L_f} \rme{(L \ot (L_B \ot L_C)_{L_{BC}})_{L_f}}{\vect{\psi}}{L_A, n_{A}} \rme{(L \ot (L_B \ot L_C)_{L_{BC}})_{L_f}}{\vect{\psi}}{L_A, n_{A'}}^{\dagger}
\label{equ:mixingtheorem1}
\end{multline}

Although this result was derived for the $^3P_0$ model ($X=1$) in Ref.\ \cite{bs08}, the derivation 
immediately generalises 
for $X=0$ (the same properties of 6-j and 9-j symbols are used) giving the same final result.

If a \emph{complete} set of degenerate intermediate states was summed over we would expect no mixing 
as the sum would just give an identity 
matrix.  However, the complete set in practice would have to span the states in the particle data 
tables\cite{PDG08} and the concept of degeneracy would be grossly violated. 
The non-trivial result of Ref.\ \cite{bs08} and here is that the above theorem requires a sum over only a 
\emph{subset} of states. The practical implication of this is that for the \emph{subset}
the approximation of degeneracy can be a good first approximation\cite{bs08}.

We consider two classes of mixing following Eq.\ \ref{mix1}:
\begin{enumerate}
\item Mixing between states of the same flavor and application to hybrid mesons.
\item Mixing between states of different flavor.  We shall draw attention to the qualitatively different way the loop sums conspire for light and heavy flavors.
\end{enumerate}

\subsection{Hybrid Mixing}
\label{sec:Mixing}

The mixing between conventional \qq~ of a given flavor in states with the same $J^{PC}$ but different 
$L$ and $S$ 
was analysed in Ref.\ \cite{bs08}.  We now apply such ideas to the mixing between a conventional \qq~ 
and a hybrid 
meson of the same $J^{PC}$ for the traditional case of $S=1$ pair creation.  Summing over a 
semi-complete set of degenerate intermediate states, mixing between hybrids 
and conventional mesons only occurs if they have the same $L$, $S$ and $J^P$.  However, such 
coincidences are not 
theoretically anticipated, at least for low-lying hybrids.  For example, in the lattice (e.g. Ref.\ 
\cite{dudek}) and models \cite{ikp,cp95,bcd} if a conventional meson has \qq~ spins in spin singlet 
(triplet), then in a gluonic hybrid with the same $J^{PC}$ the \qq~ spins will be in spin triplet (singlet). 
Thus for example, the conventional $1^{\pm\pm}$ are \qq~ triplets 
whereas their hybrid realisations are spin-singlets.  Conversely, the conventional singlets 
$0^{-+},2^{-+}$ have hybrid 
configurations with the \qq~ in spin-triplet. Eq.\ \ref{equ:mixingtheorem1} therefore forbids these 
conventional mesons and hybrids to 
mix.  A particular illustration of how the flavor-spin conspires to give destructive interference of 
hadron loops in hybrid-\qq~ mixing is the $1^{--}$ case in charmonium.

The decay to two charmed mesons involves one unit of angular momentum; this may be a relative P-wave 
between 
two charmed mesons with internal L=0 ($D,D^*$) (denoted here by $[L_B L_C;L]=[00;1]$), or be an 
internal 
L=1 ($D_0,D_1,D_2$) where an orbitally excited charmed meson is produced with a $D$ or $D^*$ in 
relative 
S-wave (denoted by $[01;0]$ which implies $[01;0]$ and $[10;0]$).

The decay of $^3S_1$, $^3D_1$ excited charmonia to $[00;1]$ is allowed but, as already illustrated 
(Ref.\ \cite{bs08}), 
these states remain orthogonal in the limit where $D,D^*$ are degenerate.  The relative couplings of 
these charmonia to
 $[01;0]$ are given in Table \ref{tablechi}\cite{bct}, where $D_{1L,1H}$ refer to the heavy quark 
basis where the light flavor quarks
 are
 respectively in $L_j =$ $p_{1/2}$ or $p_{3/2}$.  The $^3S_1 \to \sum_J D\bar{D}_J \to\ ^3D_1$ 
amplitudes are seen  to vanish here also in accord with the general theorem. This orthogonality 
survives for arbitrary mixing angle between the ``light" and ``heavy" axial mesons, 
 $D_{1L}$ and $D_{1H}$ respectively, in the mass-degenerate limit.

From the general theorem, it might appear that conventional and hybrid mesons with the same $L$, $S$ 
and $J$ but different 
parity could mix.  However, we emphasise that parity conservation is enforced by the spatial matrix 
element, for example in 
limiting the combinations of mesons and partial waves ($L$) that can appear as intermediate states.

Hybrid charmonia are forbidden to decay to $[00;L]$ charmed mesons in the mass degenerate symmetry 
limit because of a zero 
from the spatial matrix element\cite{cp95}.  Hence there is no mixing of hybrid with either $^3S_1$ or 
$^3D_1$ valence states 
through $[00;1]$ loops.  The first common channels are $[01;0]$: the ground state $D,D^*$ produced in 
conjunction with orbitally 
excited $D_J$ states.  Here too, Table \ref{tablechi} shows that the overlap between hybrid and 
$^3S_1$ or hybrid and $^3D_1$ 
vanish in the symmetry limit, again in accord with the general theorem.

\begin{table}[h]
\begin{tabular}{lrrl}
\hline
        & $\psi(^3S_1)$	\qquad & \qquad	$\psi(^3D_1)$ \qquad & \qquad Hybrid		\\
\hline
 $D^*D_0$             &  $-1/2$             	\qquad &	\qquad 0  \qquad & \qquad $ 1/\sqrt{6}               	
$\\

$D^*D_{1L}$  & $1/\sqrt{2}$	\qquad &\qquad	$0		$\qquad  & \qquad $1/\sqrt{3} $ \\
$D^*D_{1H}$  & $0$	\qquad &\qquad	$1/4		$\qquad  & \qquad $ -1/2\sqrt{6} $ \\
$D^*D_{2}$  & $0$	\qquad &\qquad	$1/4\sqrt{5} $\qquad  & \qquad $-\frac{1}{2}\sqrt{\frac{5}{6}}		
$ \\
\hline
 $DD_{1L}$  & $1/2$	\qquad &\qquad	$0		$\qquad  & \qquad $-1/\sqrt{6} $ \\
$DD_{1H}$  & $0$	\qquad &\qquad	$1/2\sqrt{2}		$\qquad  & \qquad $ 1/2\sqrt{3} $ \\   
\hline
\end{tabular}
\caption{$\psi$ and hybrid vector decay amplitudes. Mixing amplitudes are proportional to sums over 
products of these numbers.}
\label{tablechi}
\end{table}

\subsection{Flavor Mixing: light flavors}
\label{sec:FlavorMixing}

\begin{figure}[ht]
\includegraphics[width=9cm]{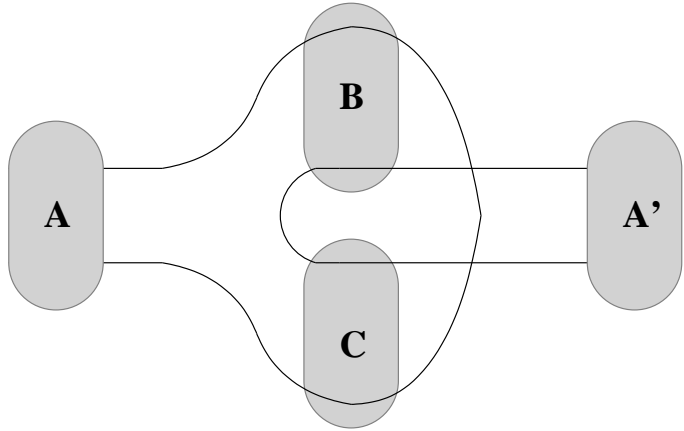}
\caption{Flavor mixing via hadronic loops}
\label{fig:LoopsflavorMixing}
\end{figure}

In Fig.\ \ref{fig:LoopsflavorMixing} we show the $Q\bar{Q} \to q\bar{q}$ loop which mixes flavors.  
Ref.\ \cite{NAT} noted that such hadronic mixing necessarily leads to breaking of the OZI rule.
Contrast how the intermediate mesons in the loop combine in this configuration relative to the 
``connected'' diagram (Fig.\ \ref{fig:LoopsMixing}) for identical flavors.  As for the ``connected'' 
diagram, this mixing diagram consists of two decay diagrams sewed together.  However, now the two 
diagrams are sewed together in a different way and this introduces an extra phase:
\begin{equation}
(-1)^{L+S_A+S_B+S_C+1}
\label{equ:extraphase}
\end{equation}
in the mixing amplitude compared to Eq.\ \ref{equ:mixfullexpression}.  
The $(-1)^{S_A+S_B+S_C+1}$ factor arises from the different spin recouplings in 
Fig.\ \ref{fig:LoopsflavorMixing} relative to Fig.\ \ref{fig:LoopsMixing}, a particular example
being Eqs.\ \ref{pseudo} and \ref{pseudoverlap}.
Integrating over the momentum flow in the two diagrams introduces the relative $(-1)^L$ factor.  

Note that this diagram can also contribute to mixing between states of the same flavor, but is expected to be suppressed, being a disconnected diagram.

If we sum over a complete subset of intermediate states we obtain the following general expression:
\begin{multline}
(m_A - m_{A'})a_{A' A} = \delta_{S_A, X} \delta_{S_{A'}, X} \delta_{J_A, J_{A'}} (-1)^L \\
\sum_{L_{BC} L_f}
\sixj{L_f}{X}{L_A}{J_A}{X}{L_{A'}}
\rme{(L \ot (L_B \ot L_C)_{L_{BC}})_{L_f}}{\vect{\psi}}{L_A, n_{A}}
\rme{(L \ot (L_B \ot L_C)_{L_{BC}})_{L_f}}{\vect{\psi}}{L_{A'}, n_{A'}}^{\dagger}
\label{equ:mixingtheorem2}
\end{multline}
This should be compared with the analogous expression for mixing between states of a given flavor in 
Eq.\ \ref{equ:mixingtheorem1}.  In the flavor mixing case, only spin-triplets mix in $S=1$
pair creation models ($X=1$), to which we shall restrict ourselves, but we can in general have $L_{A'} \neq L_A$.

We now illustrate the way that meson loops contribute to these two situations by considering $\omega$ 
and $\phi$ $1^{--}$ 
mesons coupling to strange mesons.  The relative amplitudes for $\phi$ decays to strange 
mesons are given in 
Table \ref{tablephi}. The squares of individual amplitudes give the relative rates (in the mass 
degenerate limit) for 
decays to $K\bar{K}; K\bar{K^*} + c.c; K^*\bar{K^*}$, which are in the ratio $1:4:7$. The flavor 
mixing involves the same 
amplitudes but folded together as in Fig.\ \ref{fig:LoopsflavorMixing}.  The result is that $K\bar{K}$ and $K^*\bar{K^*}$ loops have the opposite sign relative to $K\bar{K^*}$ (and its charge conjugate), having opposite G-parity (the extra phase in
 Eq.\ \ref{equ:extraphase}).  This is in accord with Refs.\ \cite{lipkin,NAT}, which argued that states of opposite G-parity interfere destructively in the flavor mixing. However, the above example shows that this is not in general a total destruction.  The ratio  of flavor mixing to total = -1/3 (i.e. - 1/12 + 4/12 - 7/12 which is just $-\left\{\begin{smallmatrix} 1 & 1 & 0 \\ 1 & 1 & 0\end{smallmatrix}\right\}$).  Thus although there is a suppression arising from the different charge conjugation/G parity intermediate states, further suppression is required from the spatial wavefunction dependence of the amplitude\cite{geiger}. 
 
 Note that this contrasts with the total orthogonality that arose for mixing between states with the 
same flavor but different 
 $L$, such as $^3S_1 - ^3D_1$, when the loops were summed corresponding to a ``connected'' topology.  This 
flavour mixing configuration (that is 
 dual to a disconnected topology) can also be considered for the $^3S_1 - ^3D_1$ situation and for 
states with the same flavor.  
 States with the same $J$, $S$, flavor and parity, but different $L$, such as $^3S_1$ and $^3D_1$ 
receive the same $1/3$ 
 suppression from flavor-spin (rather than the exact cancellation in the ``connected'' topology) but 
the overall question of 
 mixing depends on the spatial wavefunctions and the validity of the OZI rule in such multiplets.  
 
 Given that the OZI rule is 
 empirically successful for the $^3S_1$ multiplet, this suggests that there will in practice be 
suppressed mixing between the 
 $^3S_1$ and $^3D_1$ multiplets. 
 However, this is just an empirical observation of the lack of a theoretical explanation of the OZI 
rule. 
 Hence we return to this question for the case of the $\omega-\phi$ states.
 
 The same spin-flavor destructive interference factor magnitude of $1/3$ ($\left\{\begin{smallmatrix} 1 & 1 & 0 
\\ 1 & 1 & 0\end{smallmatrix}\right\}$) arises for coupling 
 $\omega-\phi$ via $K K_J$ in $L=0$ ($[01;0]$) intermediate states in the loop.  Thus although there 
is indeed a 
 flavor-spin suppression, this alone is insufficient to explain the OZI rule/lack of flavor mixing 
 in the $1^{--}$ \qq~ multiplet.  However, for $\omega-\phi$, the loops can 
couple to 
 $[00;1]$ or $[01;0]$.  The spatial amplitudes to $[00;1]$ and $[01;0]$ have a relative negative 
phase, 
 the $(-1)^L$ factor in Eq.\ \ref{equ:extraphase}.  Hence there is a destructive interference between 
these
  two multiplets in addition to the 1/3 suppression from flavor-spin.  

In general, summing over $[00;1]$ and $[01;0]$ will not give exact cancellation because of different 
counting factors ($[01;0]$ 
can be $[01;0]$ and $[10;0]$) and different spatial matrix elements.  In a specific model Geiger and 
Isgur\cite{geiger} noted that
 the spatial cancellation can be exact in an analytic limit which is near to the empirical set of 
parameters used in quark models. 
 Even away from this ideal situation, they 
 argued that summing over spatial excitations gives a destructive interference between $[00;1]$ and 
$[01;0]$ meson loops that can 
 play a significant role in minimising loop corrections to the OZI rule in $\omega-\phi$. 

For initial \qq~ states with internal $L \neq 0$ the spatial analysis become model dependent.  No 
general conclusions can be drawn other 
than if a complete set of states were summed over, in the $^3P_0$ model the OZI rule would work for 
all $J^{PC}$ states other than 
$^3P_0$.

\begin{table}[h]
\begin{tabular}{ccccc}
\hline
       
$J_{BC}$ & $\phi \to K\bar{K}$ \qquad  &  \qquad $\phi \to K\bar{K^*}$ \qquad & \qquad $\phi \to K^*\bar{K}$ \qquad &	
 \qquad $\phi \to K^*\bar{K}^*$ \\
\hline
$0$ & $1/2\sqrt{3}$ \qquad & -- & -- & \qquad 1/6 \\ 
$1$ & -- & \qquad $1/\sqrt{6}$ \qquad &  \qquad  $-1/\sqrt{6}$ \qquad & \qquad 0 \\
$2$ & -- & -- & -- & $-\sqrt{5}/3$  \\  
\hline
\end{tabular}
\caption{$\phi$ decay amplitudes ($L=1$)}
\label{tablephi}
\end{table}

The above discussion contrasts with the case of the pseudoscalar $0^{-+}$. Denoting pseudoscalar and 
vector mesons by $P,V$ respectively, and the spin projections of the vector by $\uparrow,\downarrow, 
z$, 
the OZI creation of \qq~ in spin-triplet gives in the symmetry limit the normalized state
 
 \begin{equation}
 \frac{1}{2}|V_z\bar{P} + P\bar{V_z} + V_\uparrow\bar{V_\downarrow} - V_\downarrow\bar{V_\uparrow} 
\rangle
 \label{pseudo}
 \end{equation}
 The amplitude for flavor mixing is then proportional to the overlap
 \begin{equation}
 \frac{1}{4}\langle P\bar{V_z} + V_z\bar{P} - V_\uparrow\bar{V_\downarrow}+ 
V_\downarrow\bar{V_\uparrow} 
 |V_z\bar{P} + P\bar{V_z} + V_\uparrow\bar{V_\downarrow} - V_\downarrow\bar{V_\uparrow} \rangle \equiv 
0
 \label{pseudoverlap}
 \end{equation}
 In this case the sum over amplitudes for flavor mixing vanishes in the symmetry limit, as expected 
from Eq. \ref{equ:mixingtheorem2}. This is a result 
 of the assumed \qq~ creation being spin-triplet ($^3P_0$) acting within a \qq~ state ($0^{-+}$) which 
is spin-singlet.

 We now consider the implications of these results for the case of hybrid mesons.  Hybrid $1^{--}$ 
have the \qq~
  coupled to $S=0$ and hence the mixing is strongly suppressed; hybrid vector mesons should be ideal 
flavor states 
  within the approximations employed here. While this may help distinguish a \qq~ or hybrid vector 
nonet from a 
  di-meson or tetraquark multiplet, it does not help distinguish a hybrid multiplet from a conventional \qq~ nonet.
 
Hybrids with exotic $J^{PC} = 0^{+-},1^{-+},2^{+-}$ have the \qq~ in spin-triplet.  Ref.\ 
\cite{geiger} argued that
for the conventional spin-triplet vector mesons, an essential source of the OZI rule for $ 
\omega-\phi$ is
  $[00;1]$ cancelling with $[01;0]$. 
If this is indeed a major player in realising the OZI rule, then one would not anticipate this 
cancellation
to occur for these exotic 
hybrids because the spatial matrix element to $[00;1]$ is expected to vanish.  In this case, there 
could be significant 
violation of the OZI rule, and flavor mixing in these exotic hybrids.

\subsection{Flavor mixing: heavy flavors}

When heavy flavors ($m_Q > \Lambda_{QCD}$) are involved, inter-flavor mixing is qualitatively 
different.
Consider the case of $c\bar{c}$ mixing into the $\eta$ and $n\bar{n}$ into the $\eta_c$. 
This could occur via intermediate loops involving $ D\bar{D^*}+c.c, D^*\bar{D^*}$ ($D\bar{D}$ being 
forbidden by
parity). However,  generically $c\bar{c} \to D\bar{D}$ can occur by the factorising OZI creation of 
\qq, whereas
\qq~ $\to D\bar{D}$ requires the creation of $c\bar{c}$. 
In the latter, where $2m_c > \Lambda_{QCD}$, the OZI process would require the 
color fields of force to extend over excessive distances without having 
created light $q\bar{q}$. 
This is highly improbable and the OZI process is suppressed both theoretically and 
empirically\cite{cd08}.  
A way for such decays to be triggered is if the required energy to create $c\bar{c}$ is supplied by a
hard process such as single gluon exchange. 

The \qq~ $\to q\bar{q}+g \to q\bar{c}c\bar{q}$ leaves the light flavors in color-octet.
For flavor mixing in a color-singlet meson, this will require the
subsequent annihilation of the \qq~ to mirror the former: $q\bar{c}c\bar{q} \to \bar{c}c+g \to 
\bar{c}c $.
Hence the mixing amplitude, through an intermediate loop of charmed mesons, will necessarily be of 
$O(\alpha_s^2)$
and in accord with pQCD.

We see that even for light flavors the loop mixing is in effect $O(\alpha_s^2)$ for the $0^{-+}$ case. 
In the limit where the intermediate vector and pseudoscalar mesons have equal mass ($m_V = m_P$), Eq.\ 
\ref{pseudoverlap} shows the OZI loop amplitude vanishes. A non-zero overlap follows if $m_V \neq 
m_P$.
In practice this is the case and arises due to the chromomagnetic interaction arising from one gluon 
exchange (OgE)\cite{dgg}. Phenomenologically
therefore, the non-zero mixing from OZI-generated loops is proportional to $O(\alpha_s^2)$, manifested 
by the vector-pseudoscalar mass splittings.

The mixings involving heavy flavors in spin-triplet, such as $\psi - \omega$, are required to involve 
perturbative gluons
for analogous reasons to the $0^{-+}$ case. However, charge conjugation plays a non-trivial role.
The initial state \qq~ has charge conjugation $C=-$. The initial
step \qq~ $\to q\bar{q}+g \to q\bar{c}c\bar{q}$ converts the \qq~ into $C=+$ and creates the 
$c\bar{c}$ with $C=-$.
While the latter requires only color rearrangement to map onto the $C=-$ $c\bar{c}$ final state,
the subsequent annihilation of \qq~ requires two gluons to satisfy charge conjugation, hence three 
gluons overall.
Consequently for heavy flavors in general, meson loops are dual to pQCD expectations.

\section{Discussion and Conclusion}

In the limit where all mesons in a loop belong to a degenerate subset, we have extended an existing 
theorem of Ref.\ \cite{bs08}. First, we have shown that that theorem is not restricted to \qq~ creation in $S=1$ but also applies for $S=0$. The previous result that $^3S_1$ or $^3D_1$ states of given
flavors remain orthogonal is found to apply also to hybrid mesons such that mixing of a hybrid vector 
with either $^3S_1$ or $^3D_1$ \qq~ vanishes in these circumstances. Hence within these assumptions, hybrid vector \qq~ mesons will decouple from their conventional \qq~ counterparts.

We have extended the discussion to the case of loop-induced mixing between different flavors.
For the case of heavy quarkonium the mixing with light \qq~ through loops
containing heavy-flavored mesons (for example \cc~ mixing with \qq~ via charmed mesons) is
dominated by pQCD (gluon exchange). The meson loops are suppressed by $O(\alpha_s^{[2,3]})$ as in conventional
pQCD of the form \cc~ $\to $ gluons $ \to$ \qq. 
For light flavors, by contrast, there is no connection between meson loops and pQCD, except
perhaps insofar as degeneracy can be broken by OgE.

One of the motivations of this work was to examine the circumstances under which hybrid mesons
would decouple from or be distinguishable from conventional \qq~ and other exotics. We have shown that
the vector hybrids decouple from the vector \qq~ states; there is however also the question of flavor mixing.

If for some $J^{PC}$ a hybrid nonet were flavor ideal, this would enable a clear distinction
from di-meson correlations with the same $J^{PC}$. The former would exhibit the familiar flavor 
triangle as in the
$\phi -K^* - [\omega,\rho]$ as against the inverted structure for a tetraquark/di-meson nonet, e.g. as 
in the low-lying scalars
$[f_0,a_0] - \kappa - \sigma$. To decide, a priori, which extreme is more likely requires understanding 
the origin of flavor mixing and the OZI rule
for conventional \qq.

For light flavors where \qq~ are in $S=0$, the multiplets would be ideal within the approximations 
used here.  However, these theorems can not immediately be applied to the ground-state $0^{-+}$ even within 
a naive picture where they are \qq~ states.  First of all, the loops contain the mesons themselves.  
Therefore, to be consistent some iterative procedure, such as that suggested in \cite{NAT}, would need to be used and converge. 
Second, the assumption of mass degeneracy within a subset is empirically severely violated.

For the initial \qq~ in $S=1$ we demonstrated that coupling to loops gives a destructive interference 
that is 67\% in amplitude; this quantifies a qualitative historical observation of Lipkin\cite{lipkin}. In the 
case of initial \qq~ with internal
$L=0$, namely $1^{--}$, there is further destructive interference in the spatial overlaps of meson 
loops containing $[00;1]$ and $[01;0]$ (in the $[L_B L_C;L]$ notation of Section \ref{sec:Loops}). The absolute cancellations are model dependent, as discussed by Geiger and 
Isgur\cite{geiger}.
Hence there is a qualitative understanding of the OZI rule for the $1^{--}$ nonet, but any 
quantitative description depends
on the details of strong interactions, and is currently beyond lattice QCD. 

This qualitative picture suggests that the flavor mixing for hybrid multiplets may be quite different 
from conventional \qq.
For a hybrid vector with its \qq~ coupled to $S=0$ the multiplet should be ideal, at least in $S=1$ \qq~ pair-creation dynamics. Exotic $J^{PC}$, such as $1^{-+}$, have the \qq~ coupled
to $S=1$. While the flavor-spin suppression arises here, thereby giving a tendency towards an ideal 
OZI nonet as before,
the spatial interferences are non-trivial. In a particular limit, discussed in Refs.\ \cite{ikp} and 
\cite{cp95},
it is predicted that the spatial amplitudes for hybrid decays to $[00;1]$ vanish. However, 
the amplitude
to $[01;0]$ is allowed and predicted to be the dominant decay channel. This implies that the pattern of spatial suppressions found for $\omega-\phi$ does not arise here.

In summary, the flavor-spin coupling favors ideal mixing for the $1^{-+}$ hybrid nonet but of itself 
is not sufficient to expect a situation as ideal as that found for the conventional $1^{--}$ nonet. 
The
ideal situation observed for the conventional $1^{--}$ appears also to require spatial interferences, 
which are not anticipated
to arise for the $1^{-+}$ hybrid if our current understanding of hybrid meson decays is any guide. In such a case, where 
decays to $[01;0]$
dominate, this will make it  hard to distinguish a hybrid \qq~ nonet from di-meson enhancements, 
though the
decoupling from $[00;1]$ can give a characteristic signature. Conversely, if the predicted 
decouplings
from $[00;1]$ are model-dependent artefacts, this may make an isolated hybrid harder to 
identify, but
offers the possibility that \qq~ $1^{-+}$ will be flavor-ideal and hence distinct from di-meson or 
tetraquark
correlations.

\section*{Acknowledgements}

We are indebted to T. Barnes, J. Dudek and E. Swanson for discussions.
This work is supported in part by grants from the Science \& Technology Facilities Council (UK) and in 
part authored by Jefferson Science Associates, LLC under U.S. DOE Contract No. DE-AC05-06OR23177. The 
U.S. Government retains a non-exclusive, paid-up, irrevocable, world-wide license to publish or 
reproduce this manuscript for U.S. Government purposes.



\end{document}